# Searches for Technosignatures: The State of the Profession

**Primary thematic area: Planetary Systems**, especially exobiology and the search for life beyond the Solar System. **Secondary thematic areas**: •Star and planet formation •Resolved stellar populations and their environments • Galaxy evolution

**Principal Author:  Jason T. Wright**

**Email:** astrowright@gmail.com **Phone:**(814) 863-8470

525 Davey Laboratory

Department of Astronomy & Astrophysics & Center for Habitable Worlds

Penn State University, University Park, PA 16802

| 126 endorsers | Affiliation |
| --- | --- |
| Veronica Allen | NASA GSFC/USRA |
| Julián D. Alvarado-Gómez | Center for Astrophysics | Harvard & Smithsonian |
| Daniel Angerhausen | Center for Space and Habitability, Bern University |
| Daniel Apai | University of Arizona |
| Dimitra Atri | NYU Abu Dhabi |
| Amedeo Balbi | University of Rome "Tor Vergata", Italy |
| Thomas Barclay | UMBC |
| Geert Barentsen | Bay Area Environmental Research Institute |
| Tony Beasley | National Radio Astronomy Observatory |
| Thomas Beatty | University of Arizona |
| Aida Behmard | Caltech |
| Anamaria Berea | University of Central Florida |
| Tabetha Boyajian | Louisiana State University |
| Joanna S. Bridge | University of Louisville |
| Steve Bryson | NASA Ames |
| Jeff Bytof | Independent researcher |
| Henderson Cleaves | Earth-Life Science Institute/Blue Marble Space Institute of Science |
| Knicole Colon | NASA Goddard |
| James Cordes | Cornell University |
| Keith Cowing | [Astrobiology.com](Astrobiology.com) |
| Jason Curtis | Columbia University |
| James Davenport | University of Washington |
| Paul Davies | Arizona State University |
| Julia DeMarines | University of California, Berkeley, Blue Marble Space Institute of Science |
| Kathryn Denning | York University |
| Steven Dick | 2014 Blumberg Chair (NASA/LC) |
| Chuanfei Dong | Princeton University |
| Yvan Dutil | Independent researcher |
| Peter Edmonds | Center for Astrophysics | Harvard & Smithsonian |
| Emilio Enriquez | University of California, Berkeley |
| Marshall Eubanks | Space Initiatives Inc |
| Yan Fernandez | University of Central Florida |
| Adam Frank | University of Rochester |
| Gabriel G. De la Torre | University of Cadiz. Spain |
| Vishal Gajjar | University of California, Berkeley |
| Michael Garrett | University of Manchester |
| Dawn Gelino | NASA Exoplanet Science Institute |
| Harold Geller | George Mason University |
| Daniel Giles | Illinois Institute of Technology, Adler Planetarium |
| Eliot Gillum | SETI Institute |
| Jose L. Gómez | Instituto de Astrofísica de Andalucía, Spain |
| R.J. Graham | University of Oxford |
| Claudio Grimaldi | EPFL - Lausanne - Switzerland |
| David Grinspoon | Planetary Science Institute |
| Jacob Haqq-Misra | Blue Marble Space Institute of Science |
| Greg Hellbourg | Curtin University, Australia |
| Daniel Helman | Ton Duc Thang University |
| Paul Horowitz | Harvard University |
| Andrew Howard | Caltech |
| Howard Isaacson | University of California, Berkeley |
| Albert Jackson | NASA retired |
| Tony Z. Jia | Earth-Life Science Institute/Blue Marble Space Institute of Science |
| Morgan Kainu | University of North Texas (Denton, Texas) |
| Shubham Kanodia | Pennsylvania State University |
| Steven Kawaler | Iowa State University |
| Shana Kendall | Portland State University |
| Afshin Khan | Blue Marble Space Institute of Science |
| David Kipping | Columbia University |
| Edwin Kite | University of Chicago |
| Kevin H Knuth | University at Albany (SUNY) |
| Ravi Kopparapu | NASA Goddard |
| Eric Korpela | University of California, Berkeley |
| Pauli Laine | Univeristy of Jyväskylä |
| Graham Lau | Blue Marble Space Institute of Science |
| Larry Lesyna | LXL Technology |
| Rafael Loureiro | Blue Marble Space Institute of Science - WSSU |
| Mariah MacDonald | Penn State University |
| Jean-Luc Margot | UCLA |
| Abel Mendez | PHL @ UPR Arecibo |
| Amit Mishra | University of Cape Town |
| Ian Morrison | Curtin University, Australia |
| Susan Mullally | STScI |
| Brendan Mullan | Point Park University/Blue Marble Space Institute of Science |
| Gelu Nita | New Jersey Institute of Technology |
| Karen O'Neil | Green Bank Observatory |
| Jim Pass | Astrosociology Research Institute |
| Ivan Glaucio Paulino-Lima | Blue Marble Space Institute of Science at NASA Ames Research Center |
| Elisabeth Piotelat | CNRS, France |
| Benjamin Pope | New York University |
| Sean N. Raymond | CNRS, Laboratoire d'Astrophysique de Bordeaux, France |
| George Ricker | MIT |
| Crystal S. Riley | Blue Marble Space Institute of Science |
| Paul Robertson | University of California, Irvine |
| Graça Rocha | JPL/Caltech |
| Joseph Rodriguez | CfA \| Harvard & Smithsonian |
| Lee Rosenthal | Caltech |
| Arpita Roy | California Institute of Technology |
| Daniel Rybarczyk | University of Wisconsin-Madison |
| Rafael S. de Souza | University of North Carolina at Chapel Hill |
| Shauna Sallmen | University of Wisconsin-La Crosse |
| Marcos Santander | University of Alabama |
| Caleb Scharf | Columbia University |
| Edward Schwieterman | Blue Marble Space Institute of Science |
| Douglas Seiler | SETI Institute |
| Megan Shabram | NASA Ames |
| Sofia Sheikh | Pennsylvania State University |
| Seth Shostak | SETI Institute |
| Pradipta Shrestha | Independent researcher |
| Andrew Siemion | University of California, Berkeley |
| Steinn Sigurdsson | Pennsylvania State University |
| Evan Sneed | Penn State |
| Hector Socas-Navarro | Instituto de Astrofisica de Canarias |
| David Soderblom | Space Telescope Science Institute |
| Arif Solmaz | Çağ University - Space Observation and Research Center |
| Sadasivan Subramaniam | Scientist (Retired),ADE,Bangalore |
| Akshay Suresh | Cornell University |
| Joshua Tan | CUNY/AMNH |
| Angelle Tanner | Mississippi State University |
| Jill Tarter | SETI Institute |
| Stuart F. Taylor | Participation Worldscope |
| Ryan Terrien | Carleton College |
| Jake D. Turner | Cornell University |
| Douglas Vakoch | METI International |
| Ciro Villa | RGNext |
| Lucianne Walkowicz | The Adler Planetarium |
| Sharon Xuesong Wang | Carnegie Observatories |
| Lauren Weiss | University of Hawaii, Manoa |
| Dan Werthimer | University of California, Berkeley |
| David Williams | University of California Santa Cruz |
| Joshua Winn | Princeton University |
| Ed Wishnow | UC Berkeley |
| Simon (Pete) Worden | Breakthrough Prize Foundation |
| Shelley Wright | UC San Diego |
| Erik Zackrisson | Uppsala University, Sweden |
| Michael Zanis | Seattle University |
| Philippe Zarka | Observatoire de Paris, CNRS, PSL, France |

# Abstract

The search for life in the universe is a major theme of astronomy and astrophysics for the next decade. Searches for technosignatures are complementary to searches for biosignatures, in that they offer an alternative path to discovery, and address the question of whether *complex* (i.e. technological) life exists elsewhere in the Galaxy. This approach has been endorsed in prior Decadal Reviews and National Academies reports, and yet the field still receives almost no federal support in the US.

Because of this lack of support, searches for technosignatures, precisely the part of the search of greatest public interest, suffers from a very small pool of trained practitioners. A major source of this issue is institutional inertia at NASA, which avoids the topic as a result of decades-past political grandstanding, conflation of the effort with non-scientific topics such as UFOs, and confusion regarding the scope of the term "SETI".

The Astro2020 Decadal should address this issue by making developing the field an explicit priority for the next decade. It should recommend that NASA and the NSF support training and curricular development in the field in a way that supports equity and diversity, and make explicit calls for proposals to fund searches for technosignatures.

## The Resurgence of the Search for Extraterrestrial Intelligence

The past 10 years have seen a resurgence in searches for extraterrestrial intelligence. Some of the drivers for this include:

- The Big Data revolution
- The privately-financed Breakthrough Listen Initiative
- Next generation radio, mm, and sub-mm capabilities developed by NRAO
- New developments in radio and optical/NIR instrumentation
- The advent of the WISE all-sky catalogs
- The repurposing of ground-based data collected for other reasons, e.g., radial velocity planet search data used for laser searches
- The promise of the new 30 m class observatories
- The stellar photometric revolution ushered in by *Kepler*
- The promise of LSST and other extremely high data rate projects
- The dawn of multi-messenger astronomy

The field has a high public profile, has inspired countless people to become astronomers and astrobiologists, and addresses one of the most profound questions humanity seeks concrete answers to. Success would prove complex life exists elsewhere in the universe, answering many of astrobiology's questions all at once.

Indeed, many in the public would be surprised to learn that the field is not part of the publicly-funded science portfolio, and how few scientists are actually involved in the search. The reasons for this are varied, but are ultimately social, not scientific, and will only be adequately addressed with a strong statement from the Astro2020 Decadal process.

## History of Federal Support for Searches for Technosignatures

Modern SETI—now also known as the search for technosignatures—began around 1960, with the publication of four foundational papers describing four different ways of searching for technological life in the universe: Cocconi & Morrison (1959) described a radio search strategy, Dyson (1960) argued for a search for waste heat, Bracewell (1960) rationalized a search for artifacts in the solar system, and Schwartz & Townes (1964) described a laser search strategy.

The history of SETI at NASA is covered well by Garber (1999).[1] Briefly: by the 1970's, NASA was considering investing significant resources into the field and commissioned a detailed study on a "maximal" radio SETI program, Project Cyclops. The high cost of that vision and the baseless conflation of the endeavor with UFOlogy, science fiction, and tabloid fodder led to political troubles. In 1978 Senator William Proxmire gave the NASA SETI program one of his "Golden Fleece" awards (for "fleecing" the taxpayer for a frivolous project) and successfully ended it.

Lobbying by Carl Sagan and others, along with careful bureaucratic and political persuasion, eventually restored the stature of the program, and by 1988 NASA was ready once again to begin searching in the radio despite some Congressional opposition. In October 1992 NASA launched the High Resolution Microwave Survey (HRMS), only to see it suddenly and surprisingly cut in September 1993 via a successful, last-minute amendment by Senator Richard Bryan, who proudly announced "This hopefully will be the end of Martian hunting at the taxpayer's expense", and more quietly made it known that if future NASA budgets contained a SETI program, then the entire agency budget would be at risk.

---

[1] See also Oman-Reagan (2018) for an interdisciplinary introduction to SETI and its history.

NASA learned its lesson: since then it has provided very little support for the field. To the best of our knowledge, since 1993 there have been four NASA grants and five NSF grants to support actual searches for extraterrestrial technology, totalling $4.7 million.[2] Four of these grants totalling $2.6 million supported the development of radio instrumentation and SETI@Home, and the other five totalling $2.1 million supported science and operations of radio telescopes. The first of these grants began in 2001, meaning that **the total federal expenditure on SETI in the past 18 years has been about $260,000 per year including overhead, or between 1–2 FTEs on average.**

For this reason, most SETI practitioners are effectively "hobbyists," writing papers in their spare time on how one might perform SETI searches if one were funded, or addressing theoretical issues such as "resolutions" to the Fermi Paradox or calculations via the Drake Equation.

Actual searches have thus been funded primarily via philanthropic efforts. The three largest injections of funds (>$10 million) have been for mostly for radio SETI work: by the Barney Oliver estate to support the SETI Institute; by Paul Allen to the SETI Institute to build the Allen Telescope Array; and by Yuri Milner to fund the Breakthrough Listen Initiative centered at UC Berkeley. Smaller initiatives include support from the Planetary Society for decades, primarily for optical and radio SETI work, and scattered smaller projects such as a grants via the Templeton Foundation for, for instance, waste heat SETI work. Essentially no work has been funded for solar system artifact SETI or other searches for technosignatures.

## Recent NASA Resistance to Funding Searches for Technosignatures

There is a strong perception within the technosignature search community that NASA has an institutional hostility, or at least staunch resistance, to including searches for technosignatures in its research portfolio. This perception comes from, among other things, private discussions with NASA officials, public comments by NASA officials, and explicit (and often contradictory) language in NASA funding calls excluding searches for technosignatures. See Tarter et al. (2018) for a history of some of the formal exclusionary language through the decades.

---

[2] For our list, see
https://sites.psu.edu/astrowright/2019/03/13/do-nasa-and-the-nsf-support-seti/

The community has long perceived that part of the difficulty is persistent confusion within and beyond NASA about whether "SETI" refers to a specific radio signal search rationale and technique, the SETI Institute itself, or to the broader effort to identify technological life in the universe. This led Tarter (2007) to coin the term "technosignatures" by analogy with biosignatures to refer to any sign of technological life, including communicative transmissions and other technologies. In general, "the search for technosignatures" is used synonymously with "SETI" in the SETI community (see, e.g., Wright et al. 2018).

The rebranding seems to have helped. In 2018 Congress considered requiring NASA to include searches for technosignatures (by that name) within its science portfolio as part of the House version of the NASA appropriations bill. In anticipation of this language (which did not ultimately become law) NASA hosted the "NASA Technosignatures Workshop" in September 2018 to learn about the searches for technosignatures and how it could support them, because it had little internal expertise on the topic. At the workshop NASA Deputy Associate Administrator for Research Michael New expressed his personal view that within NASA the term "SETI" is heavily associated with HRMS and radio programs, while the term "technosignatures" is "freighted with less history" and "doesn't provoke as many antibodies" at NASA.[3]

As a result of the feedback NASA received from that workshop, most of the SETI-exclusionary language was removed in the amended 2018 ROSES document, and "technosignatures" was explicitly solicited in the amended version of the ROSES 2018 XRP and Exobiology calls in early 2019. Technosignatures remains, however, somewhat homeless at NASA. For instance: not all searches for technosignatures are focused on planetary systems (some focus on other galaxies, or the Solar System, or on all-sky searches, for instance), and so many worthy projects will not be in-scope at XRP. And some of the inconsistencies noted by Tarter et al. (2018) persist: for instance, Exobiology still explicitly excludes radio and microwave technosignatures from its scope for no reason other than institutional inertia.

## Key Issue

**The Key Issue we identify in this white paper is the small size of the trained community of SETI practitioners, and the lack of federal support for both**

---

[3] See the 13:00 mark and surrounding discussion at the video record of the first talk of the NASA Technosignatures Workshop.

**searches for technosignatures and for growing the community.** This issue has led to the neglect by NASA and the NSF of one of the primary avenues for pursuing a major component of the Planetary Systems thematic area of the Astro2020 Decadal process: the search for life in the universe.

Despite a high public profile for nearly 60 years, the field is still small and many of the search strategies identified in the 1960's remain essentially unexplored, to say nothing of the many more recently identified avenues. Indeed, an ADS bibliography of all work in the field (Reyes & Wright 2019)[4] contains fewer than 2,000 entries, only 66 of which have more than 20 citations—**the h-index of the field *as a whole* is 33**.

One reason for this is that because the field has been largely funded philanthropically, much of the work has happened outside of academia and outside the aegis of the NSF, which strongly supports training as part of its scientific efforts. What SETI work has been done since 1960 has thus not been accompanied by a proportional amount of student training or formal curricular development, and has happened largely outside of academic publish-or-perish culture.

To give another metric of the academic maturity of the field: **only five people in the US have *ever* received PhDs in the physical sciences for dissertations whose primary focus was SETI**, only two of whom are astronomers today.[5] Four were trained at Harvard by Paul Horowitz, eventually building a dedicated optical SETI observatory. The fifth is Andrew Siemion, trained in radio astronomy at Berkeley by Dan Werthimer and Geoff Bower—today Dr. Siemion serves as the director of the Breakthrough Listen SETI project and as the Oliver Chair for SETI Research at the SETI Institute.

**The small size of the trained workforce in the field is a major problem for NASA and its search for life in the universe.** For instance, at the NASA Technosignatures Workshop, participants were charged with answering questions about the state of the art of the field, advances coming in the near- and medium-future to advance the field, and what role NASA could play in the field. The resulting workshop report stated:

---

[4] A "living" version of the bibliography is available on ADS via the "SETI" bibgroup
[5] Darren Leigh (1998, Harvard), Charles Coldwell (2002, Harvard), Andrew Howard (2006, Harvard, now astronomy faculty at Caltech), Andrew Siemion (2012, UC Berkeley), Curtis Mead (2013, Harvard), plus one outside the US, Ian Morrison (2017, UNSW, Australia) and a few others with one or more chapters on SETI (for instance Maggie Turnbull). There have also been PhD theses studying SETI in other fields, such as history, linguistics, anthropology, etc. (for instance by Daniel Romesberg).

*…getting good answers to the questions posed by the workshop is not a matter of asking the appropriate experts to synthesize information that already exists, it will require training and supporting scientists to do the work necessary to generate that information in the first place.* —Technosignatures Workshop Participants (2018)

Indeed, the report identified that a major initiative that NASA could undertake to determine the best ways to support future searches for technosignatures was to

*…foster a scientific and educational community in which researchers from an interdisciplinary array of fields can be introduced to the history of research into technosignatures, the current state of progress of such searches, and learn the open issues in the field and where they can contribute to the future study of and search for technosignatures.* —Technosignatures Workshop Participants (2018)

## Prior Endorsements for Searches for Technosignatures

Tarter et al. (2018) catalogs many past recommendations from National Academies reports, NASA strategy documents, and prior Decadal surveys that endorse NASA's embrace of SETI or the search for technosignatures, including the 1998, 2003, and 2008 Astrobiology Roadmaps, and the 2015 Astrobiology Strategy document. For instance, the 2010 Decadal survey stated:

*Of course, the most certain sign of extraterrestrial life would be a signal indicative of intelligence. [A radio] facility that devoted some time to the search for extraterrestrial intelligence would provide a valuable complement to the efforts suggested by the PSF report on this question. Detecting such a signal is certainly a long shot, but it may prove to be the only definitive evidence for extraterrestrial life. (p.454, Panel Reports—New Worlds, New Horizons in Astronomy & Astrophysics)*

Such language has not been effective. **Perversely, such vaguely positive but non-specific language past in Decadal surveys has been deployed by NASA as a reason *not* to fund SETI.** To wit: in a rebuttal to a (very positive) 2019 audit of the SETI Institute by the Office of the Inspector General[6] that lightly chided NASA for not funding searches for technosignatures, NASA wrote

---

[6] NASA's rebuttal is in Appendix C of the audit, [Report No. IG-19-011](#).

*…[A]lthough the 2001 and 2010 Astrophysics Decadal Surveys do express support for the development of SETI technology and approaches as well as looking "for signals produced by technologically advanced entities elsewhere in our galaxy," neither recommends NASA funding for this research, and in fact no recommendations for NASA investments in SETI research have been made in a decadal survey since 1991. It is incorrect to imply…that prior decadal surveys generally support searches for technosignatures.*

**It is thus unlikely NASA will significantly fund searches for technosignatures without *explicit* instruction to do so from the Astro2020 Decadal survey.**

On the NSF side, the ngVLA has embraced SETI as a key science goal (Croft et al. 2018). The Astro2020 Decadal should endorse this embrace and recommend a similar stance across the NSF science portfolio.

# Strategic Plan

It is time for NASA to move on from its 1990's aversion to SETI and nurture a new generation of astrobiologists that will lead the search for technosignatures into the future. The Astro2020 Decadal survey should explicitly recommend two paths to making searches for technosignatures part of the national research portfolio: funding actual searches, and training the next generation of scientists in the field.

### 1) Funding searches for technosignatures

The best way to develop SETI as a field is to actually practice it, including the development of SETI-specific instrumentation, and innovative uses of new astronomical facilities. The overall philosophy of modern SETI efforts is described in the Astro2020 science white paper "Searches for Technosignatures in Astronomy and Astrophysics" (Wright 2019a). There are many avenues ripe for investment, including many described in other Astro2020 science white papers:

1. Radio technosignatures (Margot et al. 2019)
2. Continuous and pulsed laser emission
3. Thermal infrared technosignatures (Wright 2019b)
4. Megastructures in transit (Wright 2019c)
5. Applications of data science (Berea 2019)
6. Scientific implications of (non-)detection (Haqq-Misra at al. 2019)
7. Observing the Earth as a communicating exoplanet (DeMarines et al. 2019)

8. Multi-messenger SETI
9. SETI in the Solar System
10. Interdisciplinary work in the humanities and social sciences, including risk communication, human and nonhuman communication, and identification of artificial objects among natural confounders (see, e.g., Dick 2018).

NASA and the NSF should welcome proposals to pursue all of these avenues and more as part of its search for life in the universe. The most natural places for these efforts is alongside searches for biosignatures under the umbrella of astrobiology, but since some SETI avenues will employ methods, like high resolution radio astronomy, that are somewhat foreign to that field, SETI work should not be *restricted* to that domain.

But being allowed to compete for scarce science funding in external funding calls will not be enough. NASA has for decades now nurtured and funded a community of scholars in astrobiology, and today that field thrives with detailed and sophisticated roadmaps, mission support, and curricula. It will be challenging for SETI programs to compete on a level playing field against such well developed scientific programs.

**NASA should therefore provide searches for technosignatures with the same support that it long ago began providing searches for biosignatures.** This should include workshops, special funding calls, and partnerships with academic and private research centers in addition to the traditional routes such as XRP and Exobiology. The NASA Astrobiology Institutes have reorganized, but the new Astrobiology Research Coordination Networks, especially that related to Life Detection, are natural places to embrace the search for technosignatures.

### 2) Supporting training explicitly through NSF and NASA

Today, astrobiology is an academic discipline. One can earn a PhD in the field (for instance at the University of Washington or at Penn State University), read textbooks in the field, and find it represented in graduate and undergraduate course catalogs.

But rarely in that graduate curriculum is the search for *technological* life represented. Today, there are only two regularly scheduled graduate courses in SETI, anywhere.[7] This lack of training is a large part of the reason the SETI community is so small.

---

[7] To our knowledge, the only two regular graduate courses are [UCLA's course](#), developed by Jean-Luc Margot, and [Penn State's course](#), developed by Jason Wright.

The NSF and NASA both support workforce development through undergraduate, graduate, and postdoctoral fellowships, early career awards, workshops, conferences, and symposia. **Both agencies should prioritize developing and nurturing SETI through these means.**

**Having the NSF and NASA guide the development of a SETI workforce will also ensure it is a diverse and equitable community.** The field's current demographics are not very diverse, but its small size of the field means that as it grows its demographics will quickly become dominated by its newest members. Growing the field in a deliberate and inclusive way will thus have a prompt and positive impact.

## Concluding Recommendations

**It is essential for the health of the field and for the search for life in the universe that the Astro2020 Decadal explicitly identify searches for technosignatures as a priority of NASA and the NSF for the next decade.**

The recent explicit declarations that technosignatures are in-scope in the XRP and Exobiology calls are welcome improvements. **This change should be endorsed by the Decadal survey** and recommend to be a persistent feature in future ROSES documents, **with the caveat that it should be broadened to include *all* well-justified technosignatures** (e.g. radio work should not continue to be arbitrarily excluded from the Exobiology call).

**The Decadal survey should emphasize that technosignatures is a broad field, and that projects involving aspects of predicting, modeling, or searching for technosignatures across a range of astrophysics should be explicitly encouraged to compete for funding in the most relevant calls across the federal science portfolio**, including the NSF CAREER and AAG programs and any future SPG programs, NASA graduate and postdoctoral fellowship program, NASA and NOAO telescope guest observer programs, and all NASA grant programs.

**But even more importantly, the Decadal should *explicitly* identify the growth and academic maturation of SETI as a scientific priority for the next decade**, and as one that is an essential component of the search for life elsewhere in the universe.